\documentclass[10pt,conference]{IEEEtran}
\IEEEoverridecommandlockouts
\usepackage{cite}
\usepackage{amsmath,amssymb,amsfonts}
\usepackage{algorithmic}
\usepackage{graphicx}
\usepackage{textcomp}
\usepackage{enumitem}
\usepackage{multirow}
\usepackage{amsmath}    
\usepackage{hyperref}   
\usepackage{diagbox}    

\usepackage{color, soul}
\usepackage[table,x11names]{xcolor}
\def\BibTeX{{\rm B\kern-.05em{\sc i\kern-.025em b}\kern-.08em
    T\kern-.1667em\lower.7ex\hbox{E}\kern-.125emX}}
\usepackage{tcolorbox}  

\soulregister\cite7     
\soulregister\ref7
\soulregister\pageref7

\begin{document}

\title{Data Quality for Software Vulnerability Datasets}

\author{\IEEEauthorblockN{%
    Roland Croft\IEEEauthorrefmark{1}\IEEEauthorrefmark{2}, 
    M. Ali Babar\IEEEauthorrefmark{1}\IEEEauthorrefmark{2}, 
    M. Mehdi Kholoosi\IEEEauthorrefmark{1}\IEEEauthorrefmark{2}%
    }%
    \IEEEauthorblockA{\IEEEauthorrefmark{1} School of Computer Science, CREST, University of Adelaide, Australia, \{firstname.lastname\}@adelaide.edu.au}%
    \IEEEauthorblockA{\IEEEauthorrefmark{2} Cyber Security Cooperative Research Centre, Australia}%
}

\maketitle

\begin{abstract}
The use of learning-based techniques to achieve automated software vulnerability detection has been of longstanding interest within the software security domain. These data-driven solutions are enabled by large software vulnerability datasets used for training and benchmarking. However, we observe that the quality of the data powering these solutions is currently ill-considered, hindering the reliability and value of produced outcomes. Whilst awareness of software vulnerability data preparation challenges is growing, there has been little investigation into the potential negative impacts of software vulnerability data quality. For instance, we lack confirmation that vulnerability labels are correct or consistent. Our study seeks to address such shortcomings by inspecting five inherent data quality attributes for four state-of-the-art software vulnerability datasets and the subsequent impacts that issues can have on software vulnerability prediction models. Surprisingly, we found that all the analyzed datasets exhibit some data quality problems. In particular, we found 20-71\% of vulnerability labels to be inaccurate in real-world datasets, and 17-99\% of data points were duplicated. We observed that these issues could cause significant impacts on downstream models, either preventing effective model training or inflating benchmark performance. We advocate for the need to overcome such challenges. Our findings will enable better consideration and assessment of software vulnerability data quality in the future. 
\end{abstract}


\section{Introduction}
\label{sec:introduction}

Software vulnerability detection is a vital task for achieving secure software systems \cite{mcgraw2004software}. However, traditional techniques for detecting vulnerabilities (e.g., rule-based methods) struggle in terms of scalability and false positive rates \cite{shahriar2012mitigating}. Hence, many researchers have been motivated to leverage the technical advancements of Artificial Intelligence (AI) and Machine Learning (ML) to support automatic software vulnerability detection \cite{lin2020software}. Recent studies have reported great success in this direction \cite{li2018vuldeepecker,li2021sysevr,li2021vulnerability,fu2022linevul,hin2022linevd}, with performance that surpasses traditional approaches \cite{croft2021empirical}. We refer to these learning-based techniques as Software Vulnerability Prediction (SVP). Like any data-driven task, SVP is highly data-dependent. In order to learn complex features of vulnerabilities, we require large code datasets that have been labelled as either vulnerable or non-vulnerable.

Nonetheless, software vulnerability data collection is not a trivial task \cite{croft2022data}. Labelled examples of software vulnerabilities are difficult to obtain in the real-world, as they are scarce \cite{zimmermann2010searching}, poorly documented \cite{zhou2021finding}, and limited to reported vulnerabilities \cite{croft2022noisy}. Consequently, many researchers have conducted labourious work constructing large-scale software vulnerability datasets \cite{fan2020ac,zhou2019devign,zheng2021d2a,nong2022generating}. However, we found that relatively little counterpart work has been conducted to understand the software vulnerability data quality. Whilst data quantity is important, an effective machine learning system requires adequate data quality \cite{vogelsang2019requirements}. Despite the increasing realisation of software vulnerability data preparation challenges \cite{croft2022data}, there has been relatively little effort made to provide a systematic understanding of how these challenges can potentially impact data quality, and subsequently affect the reliability of downstream software vulnerability analysis. 

A lack of understanding of data quality leads to critical barriers to advance software assurance against vulnerabilities. Data quality is an integral component of any data-driven system: \textit{garbage in, garbage out} \cite{sanders2017garbage}. Certain data biases or misinformation can make benchmark performance results misleading \cite{arp2022and,kang2022detecting,shi2022we}. This can cause models to fail to generalise to real-world scenarios \cite{he2022distribution,nong2022generating,chakraborty2021deep}, if they have not been trained with fair and realistic data. 

Thus, we set out to understand the nature of data quality for software vulnerability datasets. To achieve this, we focused on inherent data quality attributes that are intrinsic to the data itself. Table \ref{tab:characteristics} presents the five data quality attributes that we systematically analyse: accuracy, uniqueness, consistency, completeness, and currentness. For each attribute, we provide a measurement of its prevalence in existing datasets and analysis into the causes of observed issues. Our findings revealed that even state-of-the-art datasets exhibit considerable data quality challenges. As expected, we found these issues to cause significant negative impacts on both training and benchmarking of state-of-the-art SVP models. Our findings have substantial implications: 

\begin{table*}[t]
  \centering
  \caption{Inherent data quality attributes defined by ISO/IEC 25012 \cite{isoiec}.}
  \label{tab:characteristics}
  \begin{tabular}{|p{1.4cm}|p{13.2cm}|p{2.4cm}|}
    \hline
    \textbf{Attribute} & \textbf{Definition} & \textbf{Interpretation}\\
    \hline
    \textit{Accuracy} & The degree to which the data has attributes that correctly represent the true value of the intended attribute of a concept or event. & Correct labelling.\\
    \hline
    \textit{Uniqueness} & The degree to which there is no duplication in records. & No duplicate values.\\
    \hline
    \textit{Consistency} & The degree to which data has attributes that are free from contradiction and are coherent with other data. & Consistent labelling.\\
    \hline
    \textit{Completeness} & The degree to which subject data associated with an entity has values for all expected attributes and related instances. & No missing values.\\
    \hline
    \textit{Currentness} & The degree to which data has attributes that are of the right age. & Not obsolete data.\\
    \hline
\end{tabular}
\end{table*}

\begin{itemize}
    \item \textit{Data quality issues may constrain the patterns able to be learnt by SVP models.} We found that real-world datasets can face substantial issues for label correctness of the vulnerable class. Approximately 20-71\% of vulnerability labels were inaccurate. Furthermore, up to 47\% of labels were inconsistent. These issues cause models to learn false or insufficient patterns. 
    \item \textit{Software vulnerability benchmark datasets may lead to inflated performance.} A few datasets exhibited large data duplication rates, between 17-99\%. Hence, data leakage causes models to report inflated performance using standard test setups. Evaluation performance decreased by up to 82\% after removing such duplicates. 
\end{itemize}

To achieve the most reliable results from learning-based software vulnerability analytics, we must consider and address the issue of data quality. Whilst some of the observed quality issues can be solved easily through rule-based detection and removal, others cannot be remediated in this manner. As a community, we must focus on developing knowledge and tools for constructing high \textit{quality} software vulnerability datasets. Our contributions are twofold: 

\begin{enumerate}
    \item We provide understanding of the nature and causes of observed data quality issues in software vulnerability datasets. We also demonstrate the corresponding impact of these issues for software vulnerability prediction. Such investigation highlights the problem of data quality and the need to mitigate these challenges. Furthermore, our insights help overcome data quality issues, for which we have provided directions in the discussion. 
    \item We propose and conduct methods for measurement of data quality for software vulnerability datasets. These efforts can enable practitioners to consider and perform data quality assessment. We have provided a reproduction package of this study to assist with such efforts \cite{reproduction_package}. 
\end{enumerate}


\section{Background and Motivation}
\label{sec:background}

\subsection{Software Vulnerability Data Preparation}
\label{sec:data_preparation}

Software Vulnerability Prediction (SVP) models use program analysis techniques to learn software vulnerability patterns automatically from historical examples \cite{ghaffarian2017software,hanif2021rise}. Due to the unstructured nature of source code, researchers have found the most success using Deep Learning (DL) techniques that learn from program syntax and semantics \cite{lin2020software,chakraborty2021deep,fu2022linevul,hin2022linevd}. 

Hence, SVP is a data-hungry process \cite{zheng2021d2a}: the models require a large training dataset of annotated code modules, labelled as vulnerable or non-vulnerable. However, acquiring a reliable vulnerability label source is a non-trivial task; there is no oracle that can unfailingly prove the existence or absence of vulnerabilities from a codebase \cite{weinberg2008perfect}. Thus, researchers have relied on a variety of label sources to account for different shortcomings. Following prior analysis \cite{croft2022data,chakraborty2021deep}, we outline four main label categories: 

\begin{itemize}
    \item \textit{Security Vendor Provided.} Security vendors maintain vulnerability databases that aggregate information from various advisories. This provides a standardised collection of disclosed vulnerabilities. Examples include the National Vulnerability Database (NVD) \cite{NVD}, or the Snyk Vulnerability Database \cite{Snyk}. Vulnerability records often provide links to patches, which can then be traced to identify real-world source code and vulnerabilities. 
    
    \item \textit{Developer Provided.} Vulnerability databases may not properly document all vulnerabilities of a project \cite{anwar2021cleaning}. Hence, researchers may collect vulnerability fixing commits directly from developers via the development history or via a project's issue tracking systems. However, this method requires additional effort to search development artefacts for security-related defects. 
    
    \item \textit{Tool Created.} Developer or security vendor-provided labels have a major limitation of only collecting \textit{reported} vulnerabilities, which severely limits the number of examples that can be collected. In reality, vulnerabilities can remain latent or undetected \cite{jimenez2019importance}, which limits the dataset size and adds considerable label noise to the modules. To circumvent this, some researchers have utilised static security analysis tools to automatically produce labels for the source code \cite{moshtari2013using,russell2018automated,zheng2021d2a}. This process relies heavily on the accuracy of the static analyser used, which is a source of contention \cite{zhou2019devign}. 
    
    \item \textit{Synthetically Created.} Finally, to bypass the limitations of other label sources, vulnerable code examples and annotations can be created artificially from known vulnerable patterns. Synthetically producing entries ensures label correctness at the cost of source code diversity \cite{chakraborty2021deep}. 
\end{itemize}

Despite these caveats, researchers have faced data preparation challenges regardless of the selected dataset \cite{croft2022data}. Whilst state-of-the-art SVP models report good performance on benchmark datasets, performance only measures the ability of a model to fit a particular dataset. A good performance value does not guarantee that a model will generalise to real-world scenarios \cite{zhang2020machine}. Hence, data quality issues will hinder the reliability and trustworthiness of the outcomes. For instance, previous studies \cite{croft2022noisy,jimenez2019importance} have highlighted inflated performance due to inaccurate labelling mechanisms for non-vulnerable modules. Consequently, the industry value and adoption of SVP models is uncertain \cite{morrison2015challenges,sotgiu2022explainability}. Our study seeks to shed light on the state of software vulnerability data quality so that we better understand the reliability and trustworthiness of the reported outcomes that use these datasets. 

\subsection{Data Quality in Software Engineering Research}
Training data is an integral component of ML systems that  heavily influences the produced models. Unlike conventional software systems, ML systems exhibit both system and data requirements \cite{vogelsang2019requirements}. As a result, data quality is becoming an essential component of AI-based Software Engineering research \cite{liebchen2016data}. Software engineering data and artefacts are often noisy as they are usually collected \textit{post-hoc} via mining software repositories \cite{bosu2013taxonomy}. The data and labels are not generated explicitly for the purposes of research. Hence, software engineering data has been found to exhibit issues with data accuracy, relevance, and provenance \cite{bosu2013taxonomy}. 

Existing studies that investigate data quality characteristics of software engineering datasets are currently non-systematic and limited. Data quality can be defined by a large range of dimensions, like those defined in Table \ref{tab:characteristics}. Existing studies often limit their analysis to semantic or syntactic data accuracy and noise \cite{kim2011dealing,herzig2013s,shepperd2013data,wu2021data,shi2022we,sun2022importance}. Similarly, Jimenez et al. \cite{jimenez2019importance} considered label accuracy within software vulnerability datasets. However, this approach fails to provide a complete picture. To make informed data decisions, there is a need for a systematised and objective investigation of data quality in the software engineering domain. Croft et al. \cite{croft2022data} conducted a systematic literature review of the data quality issues considered by SVP researchers. Whilst this work provides a systematised view, the observations are unsubstantiated with respect to actual software vulnerability datasets and SVP models. Our work purports to perform quantitative analysis of data quality within software vulnerability datasets and explicitly show its impacts on SVP models.

Additionally, researchers have recently begun constructing automated cleaning frameworks to reduce the observed data noise issues and ensure correctness \cite{shi2022we,sun2022importance}. These frameworks are grounded in a deep understanding of the data quality issues afflicting the relevant datasets. We expect the findings obtained in our study to enable the creation of a cleaning framework for software vulnerability datasets. 

\begin{figure}[b]
  \centering
  \includegraphics[width=0.595\linewidth]{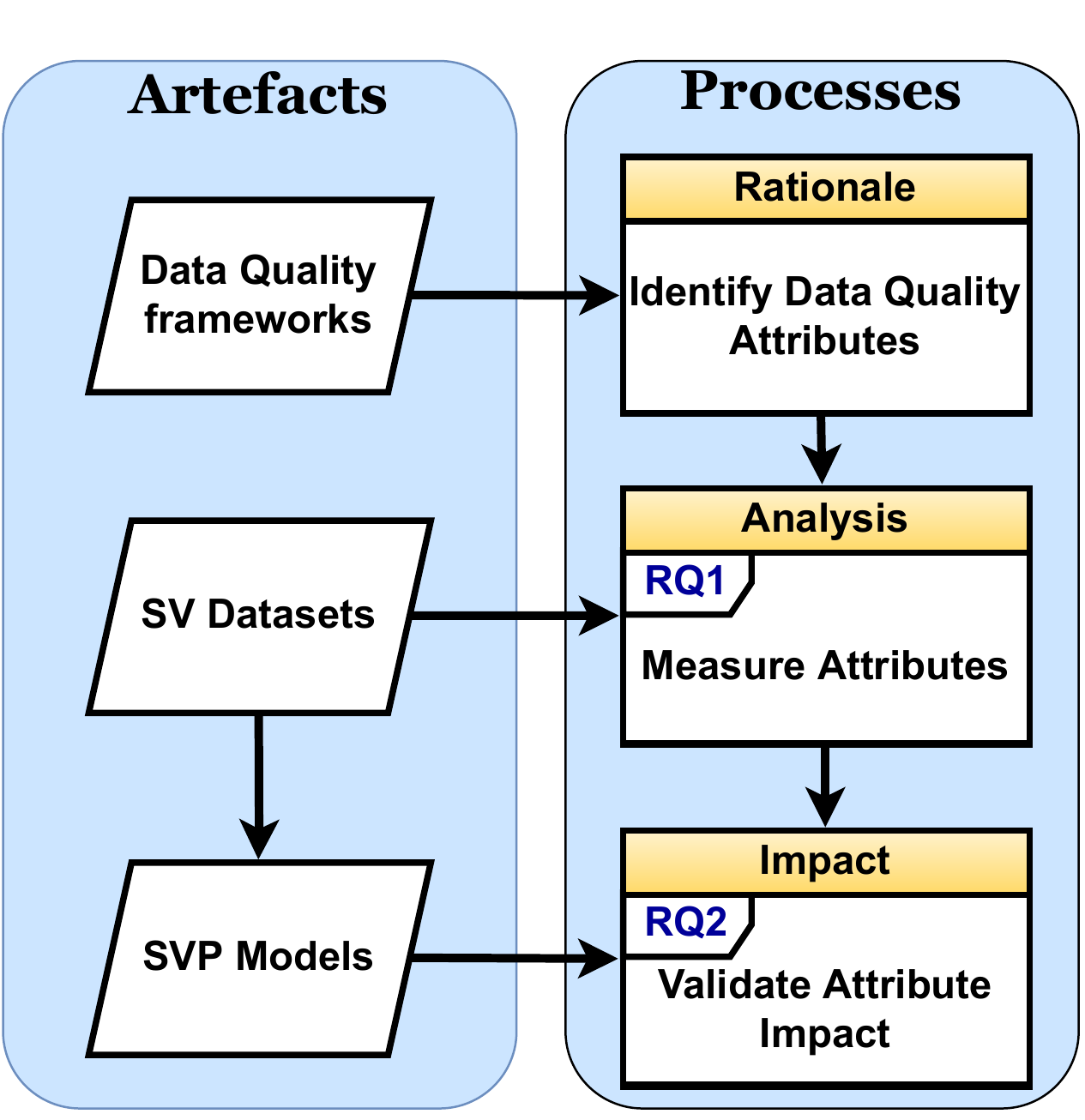}
  \caption{The overall study design.}
  \label{fig:design}
\end{figure}

\section{Study Design}
\label{sec:method}

To understand the data quality of the state-of-the-art software vulnerability datasets, we address two Research Questions (RQs) through the analysis of several data quality attributes. Figure \ref{fig:design} displays the overall workflow used to conduct this study. We describe each of the three processes in Sections \ref{sec:elicitation}, \ref{sec:measurement} and \ref{sec:validation}, respectively. 

\subsection{Research Questions}
Our investigation is guided by the following RQs: 

\begin{itemize}
    \item \textbf{RQ1: What data quality issues are present in the state-of-the-art software vulnerability datasets?} We firstly aim to inform practitioners of the state and nature of data quality for software vulnerability datasets. 
    \item \textbf{RQ2: To what extent do data quality issues impact downstream software vulnerability prediction models?} Our second aim is to verify the importance of data quality.  We attempt to demonstrate the potentially negative impact that observed data quality issues can have on the downstream tasks for software vulnerability prediction. 
\end{itemize}

\subsection{Identifying Data Quality Attributes}
\label{sec:elicitation}

The lack of existing data quality consideration for software vulnerability datasets has perhaps been caused by the lack of systematic definitions for data quality and hence measurement. It is not easy to define data quality, due to its many dimensions \cite{sidi2012data}. Not all dimensions may be relevant to a specific scenario \cite{gualo2021data}, and different organisations would value quality attributes differently \cite{vogelsang2019requirements}. For instance, not all organisations would prioritise data confidentiality. There are two main categories of data quality attributes \cite{isoiec}: \textbf{inherent data quality}, which intrinsically relates to the data itself, or \textbf{system-dependent data quality}, which extrinsically arises from external factors and requirements. For this study, we focused on purely inherent data quality attributes, as SVP models have not yet achieved widespread industrial application \cite{morrison2015challenges}. System-dependent attributes cannot be properly measured without an associated deployment context. In this sense, we focused on the data rather than how the data is used. Hence, our findings are not constrained to particular modelling techniques or features. 

To identify inherent data quality attributes, we used the standardised data quality framework ISO/IEC 25012 \cite{isoiec}. This framework has been used for data quality assessment in both the Software Engineering \cite{gualo2021data} and Machine Learning \cite{vogelsang2019requirements,nakajima2021ai} domains. ISO/IEC 25012 outlines five inherent data quality attributes: Accuracy, Consistency, Completeness, Currentness, and Credibility. We excluded the credibility attribute as it is difficult to quantify in existing software vulnerability datasets. Credibility indicates the level of \textit{trust} that we have in a dataset; the authenticity of the data source or supplier. We need to ensure that our data points are free from contamination or fake information \cite{zhang2020machine}. As datasets have been produced by peer-reviewed research or respected government organisations \cite{croft2022data}, we assume a level of trust in the data source and supplier of each dataset. Additionally, we considered a \textit{uniqueness} data dimension, due to its prevalence in existing software engineering research \cite{allamanis2019adverse}, and its importance as highlighted by previous SVP researchers \cite{chakraborty2021deep}. Table \ref{tab:characteristics} summarises the selected inherent data quality attributes for analysis. 

\subsection{Measuring Attributes}
\label{sec:measurement}

For the analysis, we collected one dataset of each label source described in Section \ref{sec:data_preparation}. To ensure the collected datasets represented the state-of-the-art appropriately, we considered datasets that were created or used by a conference or journal paper published in a high quality venue, as indicated by a CORE\footnote{\url{http://portal.core.edu.au/conf-ranks/}, \url{http://portal.core.edu.au/jnl-ranks/}} ranking of A or A*. New datasets continue to be published in order to improve on previous dataset shortcomings. We selected the most recently published datasets as of March 2022. We also chose datasets that contained appropriate metadata about how the labels were obtained. Table \ref{tab:datasets} displays our selected datasets. All four of the examined datasets provide source code for C/C++ functions.  

\begin{table}[t]
  \centering
  \caption{Selected state-of-the-art datasets for examination.}
  \label{tab:datasets}
  \begin{tabular}{cccc}
    \hline
    \textbf{Dataset} & \textbf{Label source} & \textbf{\# Functions} & \textbf{\% Vul}\\
    \hline
    Big-Vul \cite{fan2020ac} & Security vendor provided & 188,636 & 5.78 \\
    Devign \cite{zhou2019devign} & Developer provided & 27,318 & 45.61 \\
    D2A \cite{zheng2021d2a} & Tool created & 1,295,623 & 1.44 \\
    Juliet \cite{boland2012juliet} & Synthetically created & 253,002 & 36.77 \\
    \hline
\end{tabular}
\end{table}

\begin{itemize}
    \item \textbf{Big-Vul} \cite{fan2020ac} scraped software versions prior to a vulnerability fix through linked patches from the CVE Details database. Functions with lines changed in a patch were labelled as vulnerable. All remaining functions in a file touched by a commit were labelled as non-vulnerable.  
    \item \textbf{Devign} \cite{zhou2019devign} used a similar data collection method by scraping vulnerability fixes directly from GitHub commits. A keyword approach was used to separate vulnerability and non-vulnerability related commits. The vulnerability-related commits were then filtered manually to ensure accuracy. All relevant functions of each commit were collected for their respective classes.  
    \item \textbf{D2A} \cite{zheng2021d2a} collected source code by running a static analysis tool on project versions before and after bug fixing commits of six open source repositories. The vulnerable class was formed from tool warnings that disappeared in the post-fix version. The non-vulnerable class consists of the remaining tool warnings. Each data entry indicates a function containing the original vulnerability location. We retrieved all such functions to form the D2A dataset. 
    \item \textbf{Juliet} \cite{boland2012juliet} contains synthetically generated examples of programs demonstrating a variety of known vulnerable code patterns. Programs are generated automatically, based on pre-defined augmentation rules. Each program contains a vulnerable version and non-vulnerable version. Juliet was originally created to test static and dynamic security tools, but it has also been used for training SVP models. Source code data is provided in files with functions being annotated as vulnerable or not. We retrieved all functions from each annotated section of the data. 
\end{itemize}

We followed the measurement practices specified by Nakajima and Nakatani \cite{nakajima2021ai} for using the ISO/IEC 25012 framework with respect to AI training data requirements. Table \ref{tab:characteristics} describes the interpretation of each attribute. We identified the percentage of samples in a dataset that satisfy the relevant characteristics to produce an overall measurement. Hence, the measurement value for each attribute lies between 0 and 1, with 1 indicating no data quality issues are present. We formally define this measurement in Equation \ref{eq:dq}, where $N$ denotes the number of samples in a dataset and $dq(i)$ returns 1 if a data entry $i$ satisfies the relevant characteristic. Further details of the measurements for each individual attribute are provided later in Section \ref{sec:findings}.

\begin{equation}
    \label{eq:dq}
    Attribute = \sum_{i=1}^{N} \frac{dq(i)}{N}
\end{equation}

\subsection{Validating Attribute Impact}
\label{sec:validation}

Finally, to validate the impact of the observed data quality issues, we investigated the performance impacts on a state-of-the-art SVP model. Depending on the observed data quality issues, we either measured the performance change on a retrained model after mitigating data quality issues or altered the test setup to highlight the data quality characteristic of focus. We provide further details in Section \ref{sec:findings}. 

For the benchmark performance, we trained a model on each dataset, without any pre-processing of the data. However, we removed inconsistent entries for the D2A benchmark, as we were otherwise unable to produce an effective classifier for this dataset. We ran all experiments five times using random 80:10:10 training/validation/test splits unless otherwise specified, as this is a standard test setup in prior research \cite{fu2022linevul,hin2022linevd,li2021vulnerability}. 

We selected the \textit{LineVul} SVP model \cite{fu2022linevul}, as it is a recently published model that has been shown to outperform all previous baselines for both function level and line level predictions. LineVul \cite{fu2022linevul} relies on CodeBERT \cite{feng2020codebert} to obtain code feature representations that capture lexical and logical semantics. CodeBERT is a pre-trained state-of-the-art code embedding model based on the RoBERTa architecture \cite{liu2019roberta}. Similar studies have demonstrated the effectiveness of CodeBERT for SVP \cite{croft2022noisy,hin2022linevd}. LineVul generates function-level predictions using a transformer-based architecture. Although \textit{LineVul} also has the capability to localise its predictions to the line-level after performing the function-level prediction, all of the selected datasets provide labels at the function-level. Hence, we perform prediction at the function-level granularity. 

We evaluated model performance using Recall, Precision and Matthews Correlation Coefficient (MCC). We opted to use MCC as an overall indicator of performance, as its use has been recommended for similar tasks \cite{yao2020assessing}. MCC values range between -1 and 1, with 1 being the optimal value.

\section{Data Quality Analysis}
\label{sec:findings}
Table \ref{tab:measurements} displays the attribute values for each dataset. 

\begin{table}[b]
  \centering
  \caption{Measured value of each attribute for each dataset.}
  \centering
  \label{tab:measurements}
  \begin{tabular}{c|cccc}
    \hline
    \diagbox{\textbf{Attribute}}{\textbf{Dataset}} & \textbf{Big-Vul} & \textbf{Devign} & \textbf{D2A} & \textbf{Juliet} \\
    \hline
    \textbf{Accuracy*} & 0.543 & 0.800 & 0.286 & 1.000 \\
    \textbf{Uniqueness} & 0.830 & 0.899 & 0.021 & 0.163 \\
    \textbf{Consistency} & 0.999 & 0.991 & 0.531 & 0.750 \\
    \textbf{Completeness} & 0.824 & 0.944 & 0.981 & 1.000 \\
    \textbf{Currentness} & 0.761 & 0.811 & 0.844 & - \\

    \multicolumn{5}{p{7.2cm}}{\footnotesize * Based on a sample of the data.}
\end{tabular}
\end{table}

\subsection{Accuracy}
\textbf{Rationale.} Accuracy defines the correctness of the data points that comprise a dataset. This largely relates to the semantic label correctness; i.e., whether or not data points labelled as vulnerable or non-vulnerable genuinely align. It has previously been observed that non-vulnerable labels are unreliable in real-world datasets as there is no ground truth label source for this class \cite{jimenez2019importance,croft2022noisy,croft2022data}. No oracle can reliably ensure the security and absence of exploits in a given code snippet. Hence, non-vulnerable labels are usually collected simply through the absence of a vulnerable label. Thus, our analysis was constrained to the vulnerable label source. We focused our investigation on label correctness of data points labelled as vulnerable. 

\textbf{Analysis.} We determined if a label is correct via manual analysis with respect to each dataset's labelling mechanism: whether a vulnerability accurately represents the vulnerability report or static analysis tool warning that it was derived from. In this sense, we did not verify whether a vulnerability was actually exploitable, but rather whether a code snippet is functionally relevant to the reported vulnerability of each label. The following steps were taken to assess the label correctness of each entry:
\begin{enumerate}
    \item We first extracted information relating to the vulnerability and fixing commit of each dataset. All datasets provided a git fixing commit ID except Juliet. Big-Vul also provided CVE-IDs and D2A contained the static analysis tool trace. 
    \item We read the fixing commit description and other available information (i.e., the vulnerability description from NVD for Big-Vul and the static tool trace for D2A) to gain an understanding of the vulnerability and the fixing commit changes. 
    \item We then examined the changed lines in the fixing commit for the relevant function, as well as the entire function's  code to understand the context of the changed lines. Based on this code comprehension, we made an assessment as to whether the changed lines were functionally relevant to the information from the previous step. 
    \item If we did not interpret them as functionally relevant, we examined all the fixing commit changes to identify where the root changes were to understand why the flagged function was not relevant. 
    \item Afterwards, the authors discussed the labels that were in disagreement and reached a consensus. 
\end{enumerate}

To facilitate our manual review, we examined 70 random samples of each dataset (90\% confidence level +/- 10\% \cite{cochran2007sampling}). Two of the authors of this paper conducted this manual analysis independently; each of them had two to five years of software security-related experience gained in academia and industry. The two raters achieved a Cohen Kappa value of 0.627 \cite{cohen1960}, which implies moderate to strong agreement.

Our findings revealed that label inaccuracy occurred within the real-world datasets. We obtained accuracy values of 0.8 (Devign), 0.543 (Big-Vul), and 0.286 (D2A). We found no inaccuracies within the synthetic Juliet dataset, as the vulnerable cases are crafted specifically for the label rather than collected \textit{post-hoc}. Real-world labelling works by tracing a vulnerability identifier (usually a vulnerability fix or warning) to the original code snippet. The two authors who conducted the manual labelling noted their reasoning behind a label being correct or incorrect. We conducted a thematic analysis \cite{braun2006using} of the label reasoning to identify the causes of dataset inaccuracy. Table \ref{tab:acc_themes} displays the proportion of each theme. 

\begin{table}[b]
  \centering
  \caption{Types of label inaccuracy in real-world datasets.}
  \label{tab:acc_themes}
  \begin{tabular}{cccc}
    \hline
    \textbf{Dataset} & \textbf{Irrelevant} & \textbf{Cleanup} & \textbf{Inaccurate}\\
    \hline
    Big-Vul & 25\% & 28.1\% & 46.9\% \\
    Devign & 42.9\% & 21.4\% & 35.7\% \\
    D2A & 0 & 0 & 100\% \\
    \hline
\end{tabular}
\end{table}

\begin{itemize}
    \item \textbf{Irrelevant code changes.} The real-world datasets largely assume that code touched by a vulnerability fix is vulnerable code. However, a vulnerability fixing commit may not necessarily provide a patch alone. Non-functional changes, such as style changes, refactoring and code migration can confuse the data labelling process. For instance, this example fixing line\footnote{\begin{tiny}\href{https://github.com/FFmpeg/FFmpeg/commit/8b2fce0d3f5a56c40c28899c9237210ca8f9cf75\#diff-73395d3a1e02aad201d2af860c5bf0fc9cb6a68c9c711ff226eeb24ea0d409a5L400}{https://github.com/FFmpeg/FFmpeg/commit/8b2fce0d3f5a56c40c28899c9237210ca8f9cf75}\end{tiny}} simply converts a constant value to the equivalent macro. Similarly, tangled commits can implement other irrelevant changes in parallel \cite{herzig2013impact}, which will be misinterpreted as vulnerable code. 
    \item \textbf{Cleanup changes.} Vulnerability fixes can sometimes be large and disparate due to the complexity of code. Tertiary changes can be made in a commit to help better facilitate a vulnerability fix, such as adding, deleting or altering variables, functions or parameters. For example, in this fixing commit\footnote{\begin{tiny}\href{https://github.com/chromium/chromium/commit/673ce95d481ea9368c4d4d43ac756ba1d6d9e608\#diff-7f736fbfd346ea57bfcaa50d2dd642f5ef5e043625524affd878d1163e148c3eL139-L141}{https://github.com/chromium/chromium/commit/673ce95d481ea9368c4d4d43ac756ba1d6d9e608}\end{tiny}} example a vulnerability occurs for when \verb|read_only| is set as \verb|True| rather than a protected memory object. The cleanup change converts \verb|False read_only| values to a \verb|nullptr|, simply to avoid confusion. These are functional changes that relate to the vulnerability fix, so we do not consider them as irrelevant. Nonetheless, they do not indicate the location of the underlying exploitable code, and hence produce false positive labels. We call these cleanup changes, although they have also been referred to as casualty changes by Sejfia et al. \cite{sejfia2021identifying}. 
    \item \textbf{Inaccurate vulnerability fix identification.} If the labelling mechanism fails to identify a vulnerability fix, the subsequent code snippet will naturally not be a vulnerability. Datasets like Big-Vul that trace vulnerability fixes from external vulnerability reports can introduce errors into this process. For instance, we found the majority of vulnerability reports for the \textit{Chromium} project to be improperly traced as this repository is not naturally hosted via GitHub. Furthermore, datasets that attempt to identify vulnerability fixes directly from commit history (Devign and D2A) can also be rife with errors. Researchers usually attempt to identify these commits through inaccurate and unreliable keyword matching methods. Lastly, D2A uses additional help from static analysis tools to identify vulnerability fixes. These tools produce many false positive vulnerability warnings. 
\end{itemize}

Mainly, we observed tangled commits to cause problems for current real-world data labelling heuristics \cite{herzig2013impact}. Current datasets assume vulnerable code to be all code touched in a vulnerability fix, but commits are messy in practice \cite{herzig2016impact}. Similarly, vague, generic or unclear commit messages can make vulnerability fix identification difficult \cite{zhou2021finding}. In contrast, correct vulnerability labels typically stem from simple, focused and well-defined vulnerability fixing commits. 

Additionally, the datasets included samples for which we found it difficult to verify or agree upon the label. This often occurred when the location of a vulnerability falls in a grey area. For instance, should the caller of a vulnerable code snippet also be labelled as such? Herbold et al. \cite{herbold2022fine} encountered similar problems in their investigation of tangled commits. Alternatively, the label source may not contain enough information in the bug report to properly trace it. We tentatively labeled these ambiguous cases as correct. However, the software security domain should work towards clear definitions that prevent such ambiguous cases, to help with ensuring label correctness. 

Devign did not exhibit as many issues, as it is the only dataset for which the creators attempted to perform manual validation of the fixing commits. However, this accuracy assurance comes at the cost of data size. Devign is the smallest of the datasets, due to the strenuous efforts of manual validation. Nonetheless, Devign still exhibits some inaccuracies. The majority of the errors came from irrelevant changes, such as refactoring or code migration, which may imply the original authors did not check for such things. 

The accuracy for Big-Vul was lower, as many of the vulnerability fixing commits used during data extraction for this dataset were large, tangled or noisy. Most errors arose from inaccuracies in tracing the fixing commits, particularly for the Chromium project. 36\% of the vulnerable entries in Big-Vul are from the Chromium project. 

Over two-thirds of the D2A labels were inaccurate. We found that this was primarily due to the static analysis tool warnings being unreliable, as well as the vulnerability commit identifier being inaccurate. The majority of commits flagged by the D2A data extractor were not actually vulnerability fixes, as the context of the security-specific words was often misinterpreted. For instance, not all commits that contained the word ``memory'' were necessarily fixing unsafe memory operations. The majority of static tool warnings were also false positives. Static analysis tools often output an indication of the reliability of a warning, based on how confident the tool is. For example, a confident integer overflow warning would know the integer data type and variable values, whereas an unreliable report may know neither. Over 97\% of the static analysis warnings included in D2A are from the lowest reliability warning class, making them often inaccurate. However, as the static analysis tools attempt to infer the location of the vulnerability directly, there were no false positives caused by irrelevant or cleanup code changes. 

\textbf{Impact.} To evaluate the impact of inaccurate labels, we retrained each model using our manually-validated samples of each dataset as a separate holdout test set. We measured model performance when using the original labels in comparison with the manually-corrected labels. We could not measure MCC as the test set had no samples that were originally labelled as non-vulnerable. The precision decreased by 29\%, 50\% and 80\% for Devign, Big-Vul and D2A, respectively, which we confirmed to be significant using a Mann-Whitney U test \cite{mann1947test} ($p<0.05$). This was because incorrect vulnerable labels caused the models to infer incorrect patterns for this class. The models were taught vulnerable patterns that were actually non-vulnerable. Hence, in terms of model evaluation, what were previously considered true positives became false positives. Correspondingly, we found that the model recall was not significantly affected (using a Mann-Whitney U test \cite{mann1947test}) as we only uncovered label inaccuracy for the vulnerable class; the number of false negatives was unchanged. These impacts are still significant however, as they can lead to high false positive rates in models which would greatly increase inspection efforts during practical use.

\begin{tcolorbox}[right=1pt, left=1pt, top=1pt, bottom=1pt, colback=white]
    \textbf{Accuracy} is limited for some real-world datasets due to their reliance on noisy and hard-to-identify vulnerability fixing commits. Accuracy issues cause SVP models to infer the wrong patterns between classes.
\end{tcolorbox}

\subsection{Uniqueness}
\textbf{Rationale.} Uniqueness is not necessarily an intrinsic data property, as a real-world data distribution may contain duplicated samples. However, code duplication has been demonstrated to have adverse effects on trained models \cite{allamanis2019adverse}. Duplicates can introduce bias in a model towards certain samples. Inflated performance values can result when duplication occurs between the training and test sets \cite{chakraborty2021deep}. Hence, ensuring uniqueness of samples within a dataset helps models generalise towards a \textit{true} data distribution \cite{zhao2021impact}. Consequently, we treat it as an inherent attribute and decided to investigate the impacts that a lack of uniqueness would have for SVP. 

Similar or identical code fragments are defined as code clones \cite{ain2019systematic}, of which there are four main types \cite{sajnani2016sourcerercc}: 
\begin{enumerate}
    \item \textbf{Type-1:} Identical code fragments, except for differences in white-space, layout and comments. 
    \item \textbf{Type-2:} Identical code fragments, except for differences in identifier names and literal values, in addition to Type-1 clone differences. 
    \item \textbf{Type-3:} Syntactically similar code fragments that differ at the statement level. The fragments have statements added, modified and/or removed with respect to each other, in addition to Type-1 and Type-2 clone differences.
    \item \textbf{Type-4:} Syntactically dissimilar code fragments that implement the same functionality. 
\end{enumerate}

We followed standard practices and considered \textit{type-3} code clones as duplicates \cite{allamanis2019adverse}. Even functionally similar code fragments will include duplicated patterns and tokens that can adversely affect the model performance and evaluation. However, for software vulnerability datasets, slight functional changes can form the difference between a vulnerable and non-vulnerable label. A typical vulnerability fix only alters a few lines of code \cite{piantadosi2019fixing}. It is important that a model is able to capture these slight functional differences across prediction classes to avoid excessive false positive or false negative rates \cite{chakraborty2021deep}. Hence, we only considered duplicates with the same labels (vulnerable or non-vulnerable) as code clones. 


\textbf{Analysis.} An entry is not unique if it is a code clone of any other entry of the same label. To identify code clones, we reused the code duplicate detector tool produced by Allamanis \cite{allamanis2019adverse}. We lowered the minimum token count of a sample to five, as functions are smaller than the files for which this tool was originally built. This tool outputs clusters of duplicates, as there can be more than one duplicate per function. 

We observed code duplication to occur within all the datasets, but less frequently for the Big-Vul and Devign datasets. We obtained a uniqueness value of 0.830 (Big-Vul), 0.899 (Devign), 0.021 (D2A), and 0.163 (Juliet). We manually examined a sample of 30 random duplicate clusters for each dataset (74 functions for Big-Vul, 79 functions for Devign, 210 functions for Juliet, 2288 functions for D2A) to understand why duplicate code entries are present. Using thematic analysis \cite{braun2006using}, we observed three main causes of code duplication in real-world datasets:  

\begin{itemize}
    \item \textbf{Updated code.} All real-world datasets collect code from multiple versions of the same code repository in order to maximise the number of vulnerabilities observed. Across the versions, subtle functional or non-functional updates to the code introduce predominantly duplicated code snippets. For vulnerable cases, these types of duplicates can imply that the code update either failed to fix the vulnerability or introduced a new one. 
    \item \textbf{Similar function sets.} A code file may contain a suite of simple modular functions. These functions are often identical in terms of variable names, logic, and layout but have slight functional differences. For example, two functions may be implemented to start and stop a process respectively, or a set of functions may each perform a unique mathematical operation on a data flow. 
    \item \textbf{Renamed functions.} Identical functions may be duplicated and renamed for use in different files and contexts. 
\end{itemize}

\begin{figure}[b]
  \centering
  \includegraphics[width=0.8\linewidth]{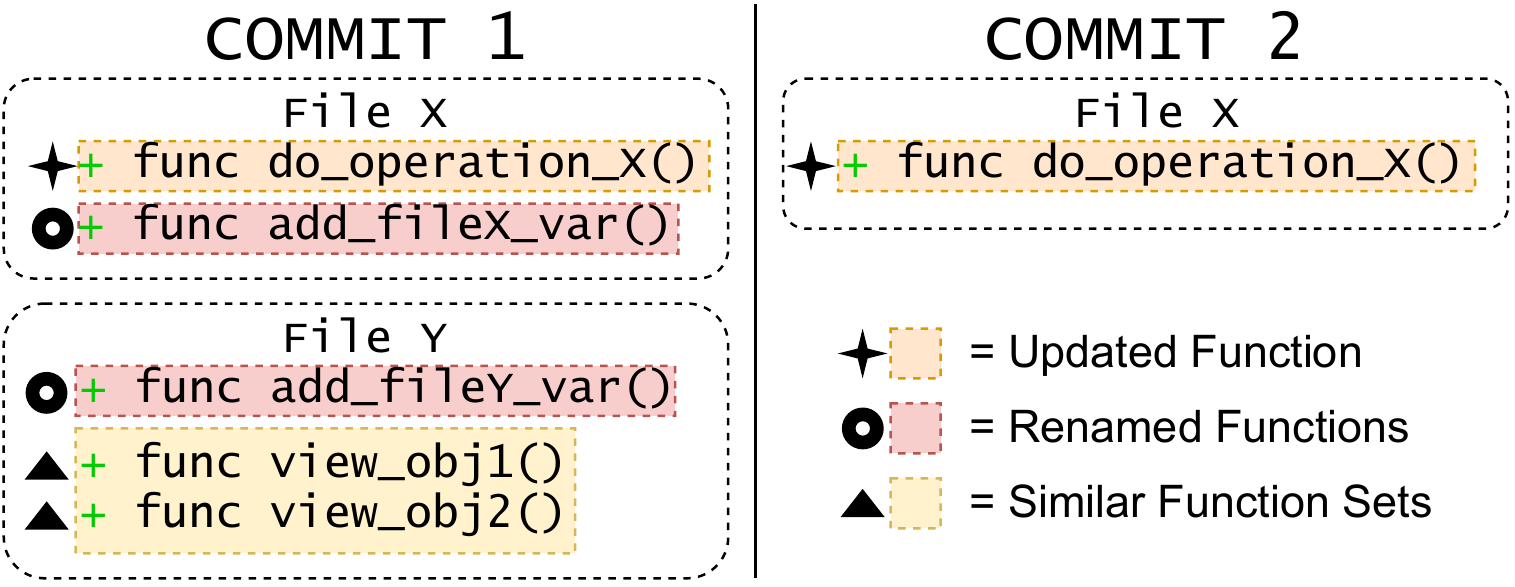}
  \caption{An example of the three main code duplicate causes.}
  \label{fig:duplicate_example}
\end{figure}

We illustrate these causes in Figure \ref{fig:duplicate_example}. These factors are inherent in source code datasets due to both the spatial and temporal repetitiveness of code in software repositories. 


We found duplication to be especially significant for D2A. Each unique function in the dataset had an average of 57 duplicates. This is because D2A produces label information at the line level, which is then abstracted to the function scope. The same function can be included multiple times if unique lines are flagged. Hence, the D2A labelling process introduces many additional exact duplicates. Over 94\% of the D2A dataset were type-1 code clones. Furthermore, two of the six repositories that comprise D2A are forks of each other (\textit{FFmpeg} and \textit{Libav}), which led to further duplication. The lack of uniqueness for D2A questions the claim of the dataset's size; there is limited information at the function level for this dataset. 

We also found a large number of duplicates in Juliet, due to the subtlety in the variance of the test cases. New test cases are produced by making slight changes to the control flow logic, internal function calls, or literal values. Furthermore, the non-vulnerable fixed statements can exhibit exact duplication due to having a constant corrected implementation. 

\begin{figure}[t]
  \centering
  \includegraphics[width=0.6\linewidth]{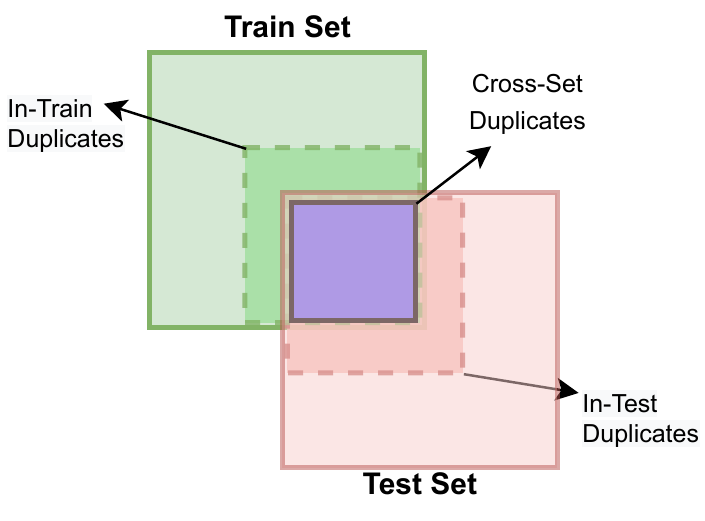}
  \caption{Types of duplicates for ML models, adapted from Allamanis \cite{allamanis2019adverse}.}
  \label{fig:dup_types}
\end{figure}

\begin{table}[t]
  \centering
  \caption{Performance impact of uniqueness issues.}
  \label{tab:uniq_impact}
  \resizebox{1.025\columnwidth}{!}{%
  \begin{tabular}{c|ccc|ccc|c}
    \hline
    \multirow{2}{*}{\textbf{Dataset}} & \multicolumn{3}{c|}{\textbf{With Duplication}} & \multicolumn{3}{c|}{\textbf{Without duplication}} & \textbf{Change}\\
    & \textbf{Precision} & \textbf{Recall}  & \textbf{MCC} & \textbf{Precision} & \textbf{Recall} & \textbf{MCC} & \textbf{(MCC)}\\
    \hline
    Big-Vul & 0.920 & 0.765& 0.830 & 0.922 & 0.762 & 0.829 & 0.0\% $\downarrow$ \\
    Devign & 0.680 & 0.428 & 0.284 & 0.651 & 0.399 & 0.244 & 13.9\% $\downarrow$ \\
    D2A & 0.961 & 0.630 & 0.774 & 0.741 & 0.049 & 0.141 & 81.7\% $\downarrow$ \\
    Juliet & 0.939 & 0.945 & 0.909 & 0.962 & 0.799 & 0.814 & 10.4\% $\downarrow$ \\
    \hline
\end{tabular}%
}
\end{table}

\textbf{Impact.} For SVP, duplicates can appear in the training set, test set, or across these two sets. Figure \ref{fig:dup_types} illustrates these duplicate types. In-train duplicates may produce model biases \cite{zhao2021impact}, but it is hard to measure these aspects via model performance \cite{allamanis2019adverse}. We focus our analysis on the impact of uniqueness for model evaluation. We split each dataset into a training, validation and test set, as specified in Section \ref{sec:validation}. We then compared the evaluation performance of the model when cross-set duplicates to the test set were either removed or kept. Allamanis \cite{allamanis2019adverse} found cross-set duplication to be the most significant type in software engineering research. 

Table \ref{tab:uniq_impact} displays the performance change for SVP models when we removed the identified duplicate entries. We observed that cross-set duplication is a significant factor for some datasets as the overall evaluation results (MCC) decreased for Devign, D2A and Juliet, which we confirmed to be significant using a Mann-Whitney U test \cite{mann1947test} ($p<0.05$). The model trained with Big-Vul data was not significantly affected. This implies that a lack of uniqueness may not always be problematic.

Duplicates can allow for data leakage in the evaluation setup \cite{allamanis2019adverse}; the models can trivially classify samples in the test set that are also duplicated in the training set, inflating the \textit{true} performance. We observed that duplication had a larger negative influence on recall rather than precision for all datasets. The removal of cross-set duplicates removed trivial samples from the test set, primarily lowering true positives. This had a larger impact on recall, due to the higher ratio of false negatives in comparison to false positives. Recall significantly decreased for Devign (7\% decrease), D2A (92\% decrease) and Juliet (15\% decrease) (confirmed using a Mann-Whitney U test \cite{mann1947test}), whereas precision actually even increased after duplicate removal for Big-Vul and Juliet. However, the overall performance (MCC) still decreased for each dataset other than Big-Vul.

\begin{tcolorbox}[right=1pt, left=1pt, top=1pt, bottom=1pt, colback=white]
    \textbf{Uniqueness} issues are present within all datasets due to the repetitive and incremental nature of code. Duplicate code snippets can potentially inflate overall evaluation performance due to data leakage. 
\end{tcolorbox}

\subsection{Consistency}
\textbf{Rationale.} Consistency denotes that data entries should not provide conflicting information. For software vulnerability datasets, this simply implies that similar code snippets should not have conflicting labels. A piece of code cannot be both vulnerable and non-vulnerable. Inconsistency can arise in software vulnerability data however, due to the multiple data streams that are used to construct a dataset \cite{croft2022investigation}. Consistency is understandably important for model training and construction, as conflicting labels confuse any AI-based model that is attempting to distinguish between two classes. 

Consistency is related to the uniqueness attribute as we again examined duplicated data. However, consistency measures duplicated entries with conflicting labels. As slight functional changes can form the functional difference between a vulnerable and non-vulnerable code snippet, we only considered type-1 code clones (exact matches). 

\textbf{Analysis.} An entry is consistent if it does not have any duplicates with conflicting labels. We observed high consistency values for Big-Vul (0.999) and Devign (0.991), but lower values for D2A (0.531) and Juliet (0.75). We manually examined a random sample of 30 inconsistent clusters to determine reasons for inconsistent vulnerability labels. We found that the causes of inconsistent labels were fairly unique to each data collection approach, which we discuss below. 

\begin{figure}[b]
  \centering
  \includegraphics[width=0.8\linewidth]{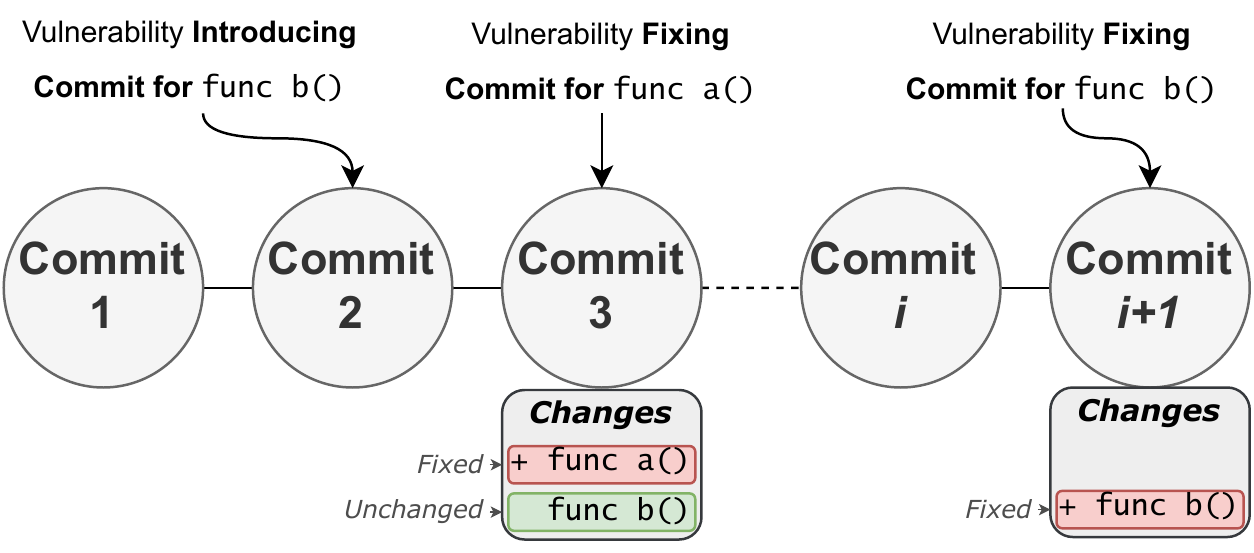}
  \caption{An example of inconsistency introduced from latent vulnerabilities. Function b is vulnerable in commit 3 until commit i+1, but it is only recorded as such for the latter.}
  \label{fig:latent}
\end{figure}

For Big-Vul, inconsistent labels were produced by latent vulnerabilities that existed within the source code. The labelling heuristic of this dataset assumes that all functions in the files of a commit that were not explicitly touched are non-vulnerable. However, these functions can actually contain vulnerabilities unknown to developers. These vulnerabilities can be reported and then collected at a later date. Figure \ref{fig:latent} illustrates this process. Although the number of inconsistent cases is relatively small, these are only the latent vulnerabilities we know about. In reality, complete knowledge of the latent vulnerabilities is unobtainable. Croft et al. \cite{croft2022noisy} observed at least twice as many latent vulnerabilities as known vulnerabilities in their dataset. 

In the Devign dataset, inconsistencies occurred due to simultaneous code branches. The vulnerability fixing commit may only be identified in one branch, leaving the same commits in other branches to be treated as non-vulnerable. This primarily occurs due to merging commits on branches, as merged commits can contain vulnerability fixes but are not described as such. Like the inconsistent labels of Big-Vul, this implies there are incorrect labels for the non-vulnerable class, as the other branch commits are improperly identified. 

The static analysis tools that inferred the labels of the D2A dataset produce an excessive number of warnings. All functions are scanned over every analysed commit during the D2A data extraction. Hence, the same function can receive the same warning from the static analysis tool over different commits. If one of the commits edits the flagged lines whereas the others do not, then inconsistent labels will be introduced. We found this occurred commonly in practice, as demonstrated by the relatively low consistency value of this dataset. 

The Juliet test cases can include tertiary functions that perform unsafe operations, e.g., writing data to a buffer. Test cases are set up like this to help test the ability of vulnerability scanning tools to track data flow across functions. Although these tertiary functions are vulnerable as they lack the necessary security checks, an exploit will only occur when specific values are passed to them. As a result, duplicate copies of these functions are contained in both the vulnerable and non-vulnerable annotated sections of this dataset. 

From the analysis, we observe that inconsistent samples primarily point to inaccuracies within the data collection processes for the non-vulnerable class. This is due to a lack of proper label sources or checks for this class; it is formed from the absence of vulnerability labels. 

\begin{table*}[t]
  \centering
  \caption{ Performance impact of consistency issues, with comparison to original data setups.}
  \label{tab:const_impact}
  \begin{tabular}{c|ccc|ccc|ccc}
    \hline
    \multirow{2}{*}{\textbf{Dataset}} & \multicolumn{3}{c|}{\textbf{All inconsistent (original)}} & \multicolumn{3}{c|}{\textbf{Consistent test set}} & \multicolumn{3}{c}{\textbf{Consistent train \& test set}}\\
    &  Precision & Recall & MCC & Precision & Recall & MCC &  Precision & Recall & MCC \\
    \hline
    Big-Vul & 0.902 & 0.774 & 0.826 & 0.919 ($\uparrow$) & 0.774 (-) & 0.835 ($\uparrow$) & 0.915 ($\uparrow$) & 0.775 ($\uparrow$) & 0.833 ($\uparrow$)\\
    Devign & 0.625 & 0.569 & 0.285 & 0.668 ($\uparrow$) & 0.500 ($\downarrow$) & 0.311 ($\uparrow$) & 0.653 ($\uparrow$) & 0.502 ($\downarrow$) & 0.289 ($\uparrow$) \\
    D2A & 0 & 0 & 0 & 0 (-) & 0 (-) & 0 (-) & 0.948 ($\uparrow$) & 0.599 ($\uparrow$) & 0.748 ($\uparrow$) \\
    Juliet & 0.937 & 0.950 & 0.910 & 0.998 ($\uparrow$) & 0.985 ($\uparrow$) & 0.987 ($\uparrow$) & 0.999 ($\uparrow$) & 0.999 ($\uparrow$) & 0.999 ($\uparrow$)\\
    \hline
\end{tabular}
\end{table*}

\textbf{Impact.} Like uniqueness, inconsistency can appear within the training set, test set, or across these two sets, as depicted in Figure \ref{fig:dup_types}. Training set inconsistency would affect the patterns learnt by the model, whereas test set inconsistency would affect model evaluation. We considered both of these aspects in our impact analysis experiments. We removed inconsistency via entries from the non-vulnerable class of inconsistent clusters, as our manual analysis found these non-vulnerable entries to be incorrect. Using the experimental setup described in Section \ref{sec:validation}, we considered three scenarios: the original case when all inconsistent examples are retained, a consistent test set in which all within-test and cross-set inconsistencies are removed but the training set remains inconsistent, and an entirely consistent dataset in which all inconsistent entries are removed. We trained and evaluated a model for each setup. 

Table \ref{tab:const_impact} displays the performance impact. We observed inconsistency to potentially have an effect on model evaluation as MCC performance increased when using consistent test sets. This is because a model will naturally make the same prediction for identical inputs, producing wrong predictions for a portion of the inconsistent entries. Hence, inconsistent samples hinder performance as the lack of distinguished labels either prevent the models from inferring important patterns or causes them to bias toward an incorrect class label. In the case of D2A, inconsistency was so prevalent that the model fails to make any correct predictions unless training with a consistent training set. We observed the model would default predictions to the most prevalent label of an inconsistent cluster; which is the non-vulnerable class in the case of D2A. Using a Mann-Whitney U test \cite{mann1947test} ($p<0.05$), we confirmed that removing inconsistency issues significantly improved performance for the most afflicted datasets (D2A and Juliet).

We observed that increased consistency has a larger positive influence on precision in comparison to recall. Recall actually even decreased when using consistent datasets for Devign (although the overall performance still increased). This is likely because inconsistent clusters more often produce false positives, due to the larger number of non-vulnerable samples in each dataset.

Performance impacts were relatively small for Big-Vul and Devign, due to the relatively small number of affected entries. We were unable to confirm whether the performance changes using these datasets were statistically significant. However, we expect that these inconsistencies actually point to larger problems in the non-vulnerable classes of these datasets. There is likely to be a much larger number of latent vulnerabilities or misclassified fixing commits, but we only observe a low number via inconsistent labels. Both Jimenez et al. \cite{jimenez2019importance} and Croft et al. \cite{croft2022noisy} found mislabelled latent vulnerabilities to impact downstream SVP models significantly. Data collection processes must be improved to ensure consistency. 

\begin{tcolorbox}[right=1pt, left=1pt, top=1pt, bottom=1pt, colback=white]
    \textbf{Consistency} issues arise due to a lack of label indicators or checks for non-vulnerable code. Whilst measured values are small; they may be an indicator of more significant problems. Consistency can be a significant issue that prevents the model from learning necessary patterns. 
\end{tcolorbox}

\subsection{Completeness}
\textbf{Rationale.} Completeness can either refer to the completeness of information within a dataset, or to the values of individual data entries. As the former requires external reference information, we focus on the latter as it is an inherent property of the data. For vulnerability datasets, source code can be missing information if the values do not contain all the code of the original function. 

\textbf{Analysis.} To detect missing information, we automatically checked for incomplete code snippets by analysing the C/C++ function syntax. We found that some code entries were missing or cut off. Overall, we observed completeness values of 0.824, 0.944, 0.981, and 1.0 for Big-Vul, Devign, D2A and Juliet, respectively. These relatively high values imply that completeness is less frequently problematic than the other data quality attributes. Missing information was only present in three of the four analyzed datasets. Table \ref{tab:completeness} displays the frequency of the truncation types present in each dataset. We have excluded Juliet because none of its entries contained missing information. 

\begin{table}[t]
  \centering
  \caption{Frequency for types of missing values in datasets.}
  \label{tab:completeness}
  \begin{tabular}{cccccc|c}
    \hline
    \multirow{2}{*}{\textbf{Dataset}} & \multicolumn{3}{c}{\textbf{Truncation}} & \multirow{2}{*}{\textbf{Empty}} & \multirow{2}{*}{\textbf{Declaration}} & \multirow{2}{*}{\textbf{Total}}\\
    & \textbf{Start} & \textbf{End} & \textbf{Both} & & & \\
    \hline
    Big-Vul & 32,973 & 133 & 140 & 0 & 0 & 33,246\\
    Devign & 814 & 265 & 9 & 0 & 0 & 1,088\\
    D2A & 0 & 0 & 0 & 10,824 & 13,300 & 24,124\\
    \hline
\end{tabular}
\end{table}

We found truncation at the start of functions to occur predominantly in the Big-Vul dataset. Return types of function definitions were truncated when they were defined over multiple lines, as function parsers commonly start on the line containing the function name. We also found a few functions in the Big-Vul and Devign dataset to be cut off prematurely, missing functional lines of code. We were unable to determine the exact cause for this truncation as we did not have access to the scripts used to produce the datasets. We hypothesise that complexities within the source code confuse the lexicographical parser being used to extract them. For instance, many of the early truncated samples contained additional curly brackets (\}) within literals. 

D2A was resilient to truncation but it contains empty missing values, for which no code was provided. These occurred when the static analysis tools flagged lines in a code file outside of any containing function. Furthermore, D2A contains 13,300 single line function declarations that do not contain any functional source code. 

\textbf{Impact.} To see the impact of missing information on SVP models, we set aside a common test set for each dataset containing no incomplete entries. We then split the remaining entries of each dataset into equal-sized halves to produce two training sets: one containing incomplete data values and the other without. The MCC performance marginally increased for the complete training sets on all datasets. However, we were unable to confirm any performance change for MCC, precision or recall as significant using a Mann-Whitney U test \cite{mann1947test} ($p>0.05$). Whilst the amount of information truncated can be of arbitrary complexity, it appears to be a relatively small part of the overall functions and occurs relatively infrequently. However, we still advise practitioners to ensure the completeness of software vulnerability data in future, as more severe issues may produce larger impact.

\begin{tcolorbox}[right=1pt, left=1pt, top=1pt, bottom=1pt, colback=white]
    \textbf{Completeness} issues can arise during data collection, but these issues are easily solvable and do not have a high impact as they cause relatively little missing information.
\end{tcolorbox}

\subsection{Currentness}
\textbf{Rationale.} Currentness aims to ensure that datasets have homogeneous temporal characteristics to their application contexts \cite{nakajima2021ai}. This is known in the machine learning domain as \textit{concept drift} \cite{gama2014survey}: a scenario in which the relationship between the input data and target variable changes over time. It is important for vulnerability datasets to stay up to date as vulnerabilities and source code have an evolving nature \cite{le2019automated,mcintosh2017fix}. We denote the date of an entry in a vulnerability dataset as the date that the vulnerability was reported via the dataset's labelling mechanism. Currentness does not relate to the synthetically created Juliet dataset.

\textbf{Analysis.} Currentness pertains to an entire dataset rather than individual data points, so we selected a standard non-contextual method for concept drift detection \cite{gama2014survey}. We used the Jensen-Shannon divergence metric \cite{fuglede2004jensen} to represent currentness, as it measures the statistical distance of the original and current data in a dataset. The formula for this metric is reported in \cite{fuglede2004jensen}. For simplicity, we denoted the original and current data as the oldest and newest half of the dataset, respectively. We represented the distribution of the vulnerability data through a Bag-Of-Tokens set. We tokenised all source code in a set using a lexicographical parser and then normalised the values based on the total frequency to obtain a probability distribution of the occurrences of each token. 

As the Jensen-Shannon divergence metric measures dissimilarity, we compute currentness as one minus this value. We obtained currentness values of 0.761 (Big-Vul), 0.811 (Devign) and 0.844 (D2A). These values are relatively high for this attribute and are unlikely to indicate concept drift.  

\begin{figure}[t]
  \centering
  \includegraphics[width=0.85\linewidth]{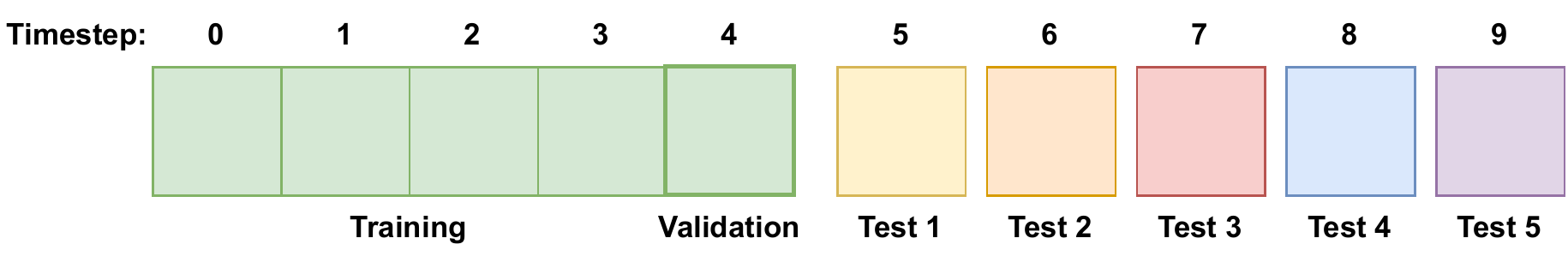}
  \caption{Currentness impact experiment setup.}
  \label{fig:currency_impact}
\end{figure}

\textbf{Impact.} We used a similar experimental setup to McIntosh et al. \cite{mcintosh2017fix} to determine whether vulnerability data is a moving target. We sorted all entries by date and then split each dataset into ten equal partitions. The four earliest partitions were used to train an SVP model, the fifth partition was used for tuning, and the remaining five were used as individual test sets. Figure \ref{fig:currency_impact} displays the experiment setup. However, using a Kendall rank correlation test \cite{kendall1938} ($p>0.05$), we observed no significant decrease in model performance for MCC, precision or recall as the time between the training and test set increased. 

\begin{tcolorbox}[right=1pt, left=1pt, top=1pt, bottom=1pt, colback=white]
    \textbf{Currentness} issues were not observed for software vulnerability datasets. They exhibit good temporal distributions of data as they are collected over a long time range.
\end{tcolorbox}

\section{Discussion}
\label{sec:discussion}

Software vulnerability datasets are particularly sensitive to data quality challenges due to the difficulties of data preparation \cite{croft2022data}. Most existing SVP studies focus on advances in modelling but often overshadow data quality. Consequently, our systematic analysis of inherent data quality attributes has revealed critical data issues afflicting the current state of software vulnerability datasets. Data preprocessing for software vulnerability data is currently cursory or inconsistent \cite{croft2022data}. There is a lack of methods to guide data cleaning efforts for SVP research. We present the following lessons learned from our quality assessment of existing datasets: 

\begin{itemize}
    \item Be wary of reusing existing datasets without first checking the data quality. 
    \item Uniqueness is poor for software vulnerability data, so avoid using evaluation setups that lead to significant duplication across the training and test set. 
    \item Inconsistently labelled data points should be removed, based on the causes of such inconsistency. 
    \item Source code entries with missing or incomplete information should be removed or amended. 
\end{itemize}

\begin{figure}[b]
  \centering
  \includegraphics[width=0.8\linewidth]{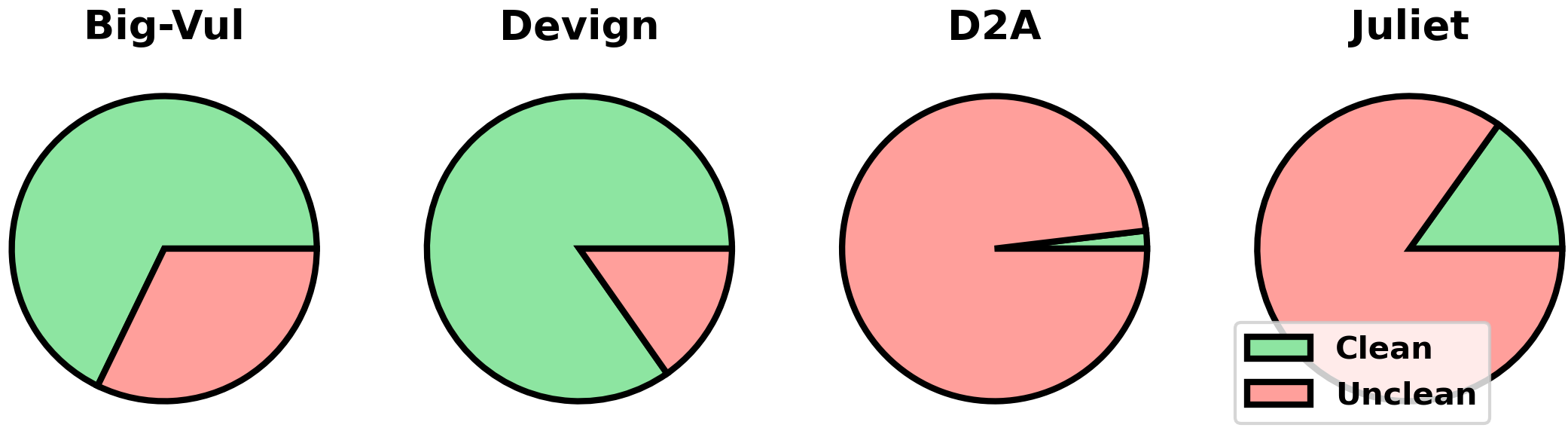}
  \caption{Ratio of clean to unclean samples in a dataset that can be automatically detected.}
  \label{fig:clean}
\end{figure}

Issues in uniqueness, consistency, and completeness can be detected with rule-based syntactic filters, as we have done in this study. Hence, we can theoretically solve these issues through exclusion of noisy samples that do not satisfy the quality attributes. However, it may not be that easy in practice as software vulnerability data is very scarce. SVP requires large datasets \cite{zheng2021d2a}, so removing noisy samples may make datasets insufficiently small. Figure \ref{fig:clean} displays the ratio of clean samples that we can automatically detect for each dataset. Furthermore, we manually observed the data accuracy issues to be severe, but there is no existing method to automatically detect such problems. Data inaccuracy could potentially decrease the number of clean entries by a further 20-71\%. 

Hence, we need to solve the underlying causes of these problems. From the findings we have obtained, we summarise the causes of the current major data quality issues below. We provide some directions for researchers to investigate, to help with overcoming these challenges: 

\begin{itemize}
    \item \textbf{Automatic data collection often leads to data inaccuracy.} We found the major cause of incorrectly labelled vulnerabilities to stem from inaccuracies in vulnerability fix identification. Either incorrect commits or line changes were selected. Substantial work has been conducted for ML-based models to identify correct vulnerability patches \cite{zhou2021finding,sawadogo2022sspcatcher}. Semantic filters or heuristics for correct vulnerability fixing lines is currently lacking. 

    \item \textbf{Source code duplication may make datasets lack diversity.} An underlying problem for data collection is a lack of sample diversity and uniqueness. Whilst we are constrained in the vulnerability samples we can collect, we have a selection choice for the non-vulnerable class. Thus, we suggest the need for development of data collection heuristics that can obtain more diverse non-vulnerable code samples. Similarly, there is a need for better synthetic data generation methods. Bug seeding has shown promising results in this regard \cite{nong2022generating}, but this technique still relies on the data quality of the real-world bugs from which the technique infers the seeds. 

    \item \textbf{Unknown vulnerabilities can introduce label inconsistency.} Label inconsistency problems arose from underlying problems for the non-vulnerable class. Big-Vul samples contained undetected vulnerabilities, and both Devign and D2A contained undocumented vulnerabilities. This is particularly problematic as we lack a label source for non-vulnerable code. Semi-supervised semantic filters have shown promise for reducing noise in non-vulnerable labels \cite{croft2022noisy,garg2022learning}. Synthetic datasets need clearly defined usage guidelines when used as training datasets. 
\end{itemize}

\section{Threats to Validity}
\label{sec:threats}

\textit{Construct Validity:} Our interpreted data quality analysis may not perfectly represent the target attributes. We have formed our analysis using standard practices from relevant domains \cite{nakajima2021ai} and existing knowledge of software vulnerability data practices \cite{croft2022data}. Requirements elicitation using domain experts would help improve these claims in future \cite{vogelsang2019requirements}. The need for manual analysis of some attributes is also a potential limitation, as it may contain bias or inaccuracies. We used two independent raters to minimize such impacts. We used CORE rankings as a criteria for our dataset selection, even though CORE journal rankings have become deprecated. We consider these ranking still sufficient, as they were only deprecated two months prior to the date of data collection.

\textit{Internal Validity:} The outcomes of our impact analysis experiments may be affected by confounding factors. We analysed each data attribute individually, so other data quality issues were present during each experiment. More work is required to examine data quality attributes cumulatively. 

\textit{External Validity:} We have constrained our analysis to four state-of-the-art datasets. Measurements are also limited to datasets that contain appropriate metadata. For instance, we were unable to investigate the ReVeal dataset \cite{chakraborty2021deep} due to this issue. Similarly, we performed impact analysis using a single SVP model. This model has been demonstrated to be state-of-the-art \cite{fu2022linevul}. Furthermore, we considered inherent data quality attributes, so the issues remain, regardless of the model.

\section{Conclusion}
\label{sec:conclusion}
We have systematically examined five data quality attributes for four state-of-the-art software vulnerability datasets, to help improve the validity and trustworthiness of downstream data-driven tasks that rely on this information. Our findings revealed that some software vulnerability datasets are prone to data quality issues, particularly in terms of data accuracy, uniqueness, and consistency. We found 20-71\% of vulnerability labels were inaccurate in real-world datasets, which altered performance up to 65\%. Furthermore, 0-47\% of the labels were inconsistent, which hindered model training completely in the most extreme circumstances. 

Data quality requires ongoing consideration and analysis. We advise future researchers and practitioners to consider data quality in more effective detail through the means that we have provided. Furthermore, we advocate the importance of data quality and the need to overcome the quality issues that we have observed. Lastly, we urge the need for additional investigation into system-dependent data quality attributes to help achieve specific operational needs. 

\section{Data Availability}
We have made our data and analysis scripts available via a reproduction package \cite{reproduction_package}.

\section*{Acknowledgment}
This work has been supported by the Cyber Security Cooperative Research Centre Limited whose activities are partially funded by the Australian Government’s Cooperative Research Centre Programme.

\bibliographystyle{IEEEtran}
\bibliography{bibfile}

\begin{thebibliography}{10}
\providecommand{\url}[1]{#1}
\csname url@samestyle\endcsname
\providecommand{\newblock}{\relax}
\providecommand{\bibinfo}[2]{#2}
\providecommand{\BIBentrySTDinterwordspacing}{\spaceskip=0pt\relax}
\providecommand{\BIBentryALTinterwordstretchfactor}{4}
\providecommand{\BIBentryALTinterwordspacing}{\spaceskip=\fontdimen2\font plus
\BIBentryALTinterwordstretchfactor\fontdimen3\font minus
  \fontdimen4\font\relax}
\providecommand{\BIBforeignlanguage}[2]{{%
\expandafter\ifx\csname l@#1\endcsname\relax
\typeout{** WARNING: IEEEtran.bst: No hyphenation pattern has been}%
\typeout{** loaded for the language `#1'. Using the pattern for}%
\typeout{** the default language instead.}%
\else
\language=\csname l@#1\endcsname
\fi
#2}}
\providecommand{\BIBdecl}{\relax}
\BIBdecl

\bibitem{mcgraw2004software}
G.~McGraw, ``Software security,'' \emph{IEEE Security \& Privacy}, vol.~2,
  no.~2, pp. 80--83, 2004.

\bibitem{shahriar2012mitigating}
H.~Shahriar and M.~Zulkernine, ``Mitigating program security vulnerabilities:
  Approaches and challenges,'' \emph{ACM Computing Surveys (CSUR)}, vol.~44,
  no.~3, pp. 1--46, 2012.

\bibitem{lin2020software}
G.~Lin, S.~Wen, Q.-L. Han, J.~Zhang, and Y.~Xiang, ``Software vulnerability
  detection using deep neural networks: a survey,'' \emph{Proceedings of the
  IEEE}, vol. 108, no.~10, pp. 1825--1848, 2020.

\bibitem{li2018vuldeepecker}
Z.~Li, D.~Zou, S.~Xu, X.~Ou, H.~Jin, S.~Wang, Z.~Deng, and Y.~Zhong,
  ``Vuldeepecker: A deep learning-based system for vulnerability detection,''
  \emph{arXiv preprint arXiv:1801.01681}, 2018.

\bibitem{li2021sysevr}
Z.~Li, D.~Zou, S.~Xu, H.~Jin, Y.~Zhu, and Z.~Chen, ``Sysevr: A framework for
  using deep learning to detect software vulnerabilities,'' \emph{IEEE
  Transactions on Dependable and Secure Computing}, 2021.

\bibitem{li2021vulnerability}
Y.~Li, S.~Wang, and T.~N. Nguyen, ``Vulnerability detection with fine-grained
  interpretations,'' in \emph{Proceedings of the 29th ACM Joint Meeting on
  European Software Engineering Conference and Symposium on the Foundations of
  Software Engineering}, 2021, pp. 292--303.

\bibitem{fu2022linevul}
M.~Fu and C.~Tantithamthavorn, ``Linevul: A transformer-based line-level
  vulnerability prediction,'' in \emph{2022 IEEE/ACM 19th International
  Conference on Mining Software Repositories (MSR)}, 2022, pp. 608--620.

\bibitem{hin2022linevd}
D.~Hin, A.~Kan, H.~Chen, and M.~A. Babar, ``Linevd: Statement-level
  vulnerability detection using graph neural networks,'' \emph{arXiv preprint
  arXiv:2203.05181}, 2022.

\bibitem{croft2021empirical}
R.~Croft, D.~Newlands, Z.~Chen, and M.~A. Babar, ``An empirical study of
  rule-based and learning-based approaches for static application security
  testing,'' in \emph{Proceedings of the 15th ACM/IEEE International Symposium
  on Empirical Software Engineering and Measurement (ESEM)}, 2021, pp. 1--12.

\bibitem{croft2022data}
R.~Croft, Y.~Xie, and M.~A. Babar, ``Data preparation for software
  vulnerability prediction: A systematic literature review,'' \emph{IEEE
  Transactions on Software Engineering}, 2022.

\bibitem{zimmermann2010searching}
T.~Zimmermann, N.~Nagappan, and L.~Williams, ``Searching for a needle in a
  haystack: Predicting security vulnerabilities for windows vista,'' in
  \emph{2010 Third International Conference on Software Testing, Verification
  and Validation}.\hskip 1em plus 0.5em minus 0.4em\relax IEEE, 2010, pp.
  421--428.

\bibitem{zhou2021finding}
J.~Zhou, M.~Pacheco, Z.~Wan, X.~Xia, D.~Lo, Y.~Wang, and A.~E. Hassan,
  ``Finding a needle in a haystack: Automated mining of silent vulnerability
  fixes,'' in \emph{2021 36th IEEE/ACM International Conference on Automated
  Software Engineering (ASE)}.\hskip 1em plus 0.5em minus 0.4em\relax IEEE,
  2021, pp. 705--716.

\bibitem{croft2022noisy}
R.~Croft, M.~A. Babar, and H.~Chen, ``Noisy label learning for security
  defects,'' \emph{arXiv preprint arXiv:2203.04468}, 2022.

\bibitem{fan2020ac}
J.~Fan, Y.~Li, S.~Wang, and T.~N. Nguyen, ``Ac/c++ code vulnerability dataset
  with code changes and cve summaries,'' in \emph{Proceedings of the 17th
  International Conference on Mining Software Repositories}, 2020, pp.
  508--512.

\bibitem{zhou2019devign}
Y.~Zhou, S.~Liu, J.~Siow, X.~Du, and Y.~Liu, ``Devign: Effective vulnerability
  identification by learning comprehensive program semantics via graph neural
  networks,'' \emph{Advances in neural information processing systems},
  vol.~32, 2019.

\bibitem{zheng2021d2a}
Y.~Zheng, S.~Pujar, B.~Lewis, L.~Buratti, E.~Epstein, B.~Yang, J.~Laredo,
  A.~Morari, and Z.~Su, ``D2a: A dataset built for ai-based vulnerability
  detection methods using differential analysis,'' in \emph{2021 IEEE/ACM 43rd
  International Conference on Software Engineering: Software Engineering in
  Practice (ICSE-SEIP)}.\hskip 1em plus 0.5em minus 0.4em\relax IEEE, 2021, pp.
  111--120.

\bibitem{nong2022generating}
Y.~Nong, Y.~Ou, M.~Pradel, F.~Chen, and H.~Cai, ``Generating realistic
  vulnerabilities via neural code editing: an empirical study,'' in
  \emph{Proceedings of the 30th ACM Joint European Software Engineering
  Conference and Symposium on the Foundations of Software Engineering}, 2022,
  pp. 1097--1109.

\bibitem{vogelsang2019requirements}
A.~Vogelsang and M.~Borg, ``Requirements engineering for machine learning:
  Perspectives from data scientists,'' in \emph{2019 IEEE 27th International
  Requirements Engineering Conference Workshops (REW)}.\hskip 1em plus 0.5em
  minus 0.4em\relax IEEE, 2019, pp. 245--251.

\bibitem{sanders2017garbage}
H.~Sanders and J.~Saxe, ``Garbage in, garbage out: how purportedly great ml
  models can be screwed up by bad data,'' \emph{Proceedings of Blackhat}, vol.
  2017, 2017.

\bibitem{arp2022and}
D.~Arp, E.~Quiring, F.~Pendlebury, A.~Warnecke, F.~Pierazzi, C.~Wressnegger,
  L.~Cavallaro, and K.~Rieck, ``Dos and don’ts of machine learning in
  computer security,'' in \emph{Proc. of the USENIX Security Symposium}, 2022.

\bibitem{kang2022detecting}
H.~J. Kang, K.~L. Aw, and D.~Lo, ``Detecting false alarms from automatic static
  analysis tools: How far are we?'' \emph{arXiv preprint arXiv:2202.05982},
  2022.

\bibitem{shi2022we}
L.~Shi, F.~Mu, X.~Chen, S.~Wang, J.~Wang, Y.~Yang, G.~Li, X.~Xia, and Q.~Wang,
  ``Are we building on the rock? on the importance of data preprocessing for
  code summarization,'' in \emph{Proceedings of the 30th ACM Joint European
  Software Engineering Conference and Symposium on the Foundations of Software
  Engineering}, 2022, pp. 107--119.

\bibitem{he2022distribution}
J.~He, L.~Beurer-Kellner, and M.~Vechev, ``On distribution shift in
  learning-based bug detectors,'' \emph{arXiv preprint arXiv:2204.10049}, 2022.

\bibitem{chakraborty2021deep}
S.~Chakraborty, R.~Krishna, Y.~Ding, and B.~Ray, ``Deep learning based
  vulnerability detection: Are we there yet,'' \emph{IEEE Transactions on
  Software Engineering}, 2021.

\bibitem{isoiec}
ISO/IEC, ``Systems and software engineering – systems and software quality
  requirements and evaluation (square) – data quality model,'' 2008.

\bibitem{reproduction_package}
\BIBentryALTinterwordspacing
R.~Croft, M.~A. Babar, and M.~Kholoosi, ``Reproduction package for ``data
  quality for software vulnerability datasets'','' 2023. [Online]. Available:
  \url{https://figshare.com/articles/software/Reproduction_Package_for_Data_Quality_for_Software_Vulnerability_Datasets_/20499924}
\BIBentrySTDinterwordspacing

\bibitem{ghaffarian2017software}
S.~M. Ghaffarian and H.~R. Shahriari, ``Software vulnerability analysis and
  discovery using machine-learning and data-mining techniques: A survey,''
  \emph{ACM Computing Surveys (CSUR)}, vol.~50, no.~4, pp. 1--36, 2017.

\bibitem{hanif2021rise}
H.~Hanif, M.~H. N.~M. Nasir, M.~F. Ab~Razak, A.~Firdaus, and N.~B. Anuar, ``The
  rise of software vulnerability: Taxonomy of software vulnerabilities
  detection and machine learning approaches,'' \emph{Journal of Network and
  Computer Applications}, p. 103009, 2021.

\bibitem{weinberg2008perfect}
G.~M. Weinberg, \emph{Perfect Software and other illusions about
  testing}.\hskip 1em plus 0.5em minus 0.4em\relax Dorset House Pub., 2008.

\bibitem{NVD}
\BIBentryALTinterwordspacing
NIST, ``\BIBforeignlanguage{en}{National vulnerability database}.'' [Online].
  Available: \url{https://nvd.nist.gov/}
\BIBentrySTDinterwordspacing

\bibitem{Snyk}
\BIBentryALTinterwordspacing
Snyk, ``\BIBforeignlanguage{en}{Snyk vulnerability database}.'' [Online].
  Available: \url{https://security.snyk.io/}
\BIBentrySTDinterwordspacing

\bibitem{anwar2021cleaning}
A.~Anwar, A.~Abusnaina, S.~Chen, F.~Li, and D.~Mohaisen, ``Cleaning the nvd:
  Comprehensive quality assessment, improvements, and analyses,'' \emph{IEEE
  Transactions on Dependable and Secure Computing}, 2021.

\bibitem{jimenez2019importance}
M.~Jimenez, R.~Rwemalika, M.~Papadakis, F.~Sarro, Y.~Le~Traon, and M.~Harman,
  ``The importance of accounting for real-world labelling when predicting
  software vulnerabilities,'' in \emph{Proceedings of the 2019 27th ACM Joint
  Meeting on European Software Engineering Conference and Symposium on the
  Foundations of Software Engineering}, 2019, pp. 695--705.

\bibitem{moshtari2013using}
S.~Moshtari, A.~Sami, and M.~Azimi, ``Using complexity metrics to improve
  software security,'' \emph{Computer Fraud \& Security}, vol. 2013, no.~5, pp.
  8--17, 2013.

\bibitem{russell2018automated}
R.~Russell, L.~Kim, L.~Hamilton, T.~Lazovich, J.~Harer, O.~Ozdemir,
  P.~Ellingwood, and M.~McConley, ``Automated vulnerability detection in source
  code using deep representation learning,'' in \emph{2018 17th IEEE
  international conference on machine learning and applications (ICMLA)}.\hskip
  1em plus 0.5em minus 0.4em\relax IEEE, 2018, pp. 757--762.

\bibitem{zhang2020machine}
J.~M. Zhang, M.~Harman, L.~Ma, and Y.~Liu, ``Machine learning testing: Survey,
  landscapes and horizons,'' \emph{IEEE Transactions on Software Engineering},
  2020.

\bibitem{morrison2015challenges}
P.~Morrison, K.~Herzig, B.~Murphy, and L.~Williams, ``Challenges with applying
  vulnerability prediction models,'' in \emph{Proceedings of the 2015 Symposium
  and Bootcamp on the Science of Security}, 2015, pp. 1--9.

\bibitem{sotgiu2022explainability}
A.~Sotgiu, M.~Pintor, and B.~Biggio, ``Explainability-based debugging of
  machine learning for vulnerability discovery,'' in \emph{Proceedings of the
  17th International Conference on Availability, Reliability and Security},
  2022, pp. 1--8.

\bibitem{liebchen2016data}
G.~Liebchen and M.~Shepperd, ``Data sets and data quality in software
  engineering: Eight years on,'' in \emph{Proceedings of the The 12th
  International Conference on Predictive Models and Data Analytics in Software
  Engineering}, 2016, pp. 1--4.

\bibitem{bosu2013taxonomy}
M.~F. Bosu and S.~G. MacDonell, ``A taxonomy of data quality challenges in
  empirical software engineering,'' in \emph{2013 22nd Australian Software
  Engineering Conference}.\hskip 1em plus 0.5em minus 0.4em\relax IEEE, 2013,
  pp. 97--106.

\bibitem{kim2011dealing}
S.~Kim, H.~Zhang, R.~Wu, and L.~Gong, ``Dealing with noise in defect
  prediction,'' in \emph{2011 33rd international conference on software
  engineering (ICSE)}.\hskip 1em plus 0.5em minus 0.4em\relax IEEE, 2011, pp.
  481--490.

\bibitem{herzig2013s}
K.~Herzig, S.~Just, and A.~Zeller, ``It's not a bug, it's a feature: how
  misclassification impacts bug prediction,'' in \emph{2013 35th international
  conference on software engineering (ICSE)}.\hskip 1em plus 0.5em minus
  0.4em\relax IEEE, 2013, pp. 392--401.

\bibitem{shepperd2013data}
M.~Shepperd, Q.~Song, Z.~Sun, and C.~Mair, ``Data quality: Some comments on the
  nasa software defect datasets,'' \emph{IEEE Transactions on Software
  Engineering}, vol.~39, no.~9, pp. 1208--1215, 2013.

\bibitem{wu2021data}
X.~Wu, W.~Zheng, X.~Xia, and D.~Lo, ``Data quality matters: A case study on
  data label correctness for security bug report prediction,'' \emph{IEEE
  Transactions on Software Engineering}, 2021.

\bibitem{sun2022importance}
Z.~Sun, L.~Li, Y.~Liu, X.~Du, and L.~Li, ``On the importance of building
  high-quality training datasets for neural code search,'' in \emph{Proceedings
  of the 44th International Conference on Software Engineering}, 2022, pp.
  1609--1620.

\bibitem{sidi2012data}
F.~Sidi, P.~H.~S. Panahy, L.~S. Affendey, M.~A. Jabar, H.~Ibrahim, and
  A.~Mustapha, ``Data quality: A survey of data quality dimensions,'' in
  \emph{2012 International Conference on Information Retrieval \& Knowledge
  Management}.\hskip 1em plus 0.5em minus 0.4em\relax IEEE, 2012, pp. 300--304.

\bibitem{gualo2021data}
F.~Gualo, M.~Rodr{\'\i}guez, J.~Verdugo, I.~Caballero, and M.~Piattini, ``Data
  quality certification using iso/iec 25012: Industrial experiences,''
  \emph{Journal of Systems and Software}, vol. 176, p. 110938, 2021.

\bibitem{nakajima2021ai}
S.~Nakajima and T.~Nakatani, ``Ai extension of square data quality model,'' in
  \emph{2021 IEEE 21st International Conference on Software Quality,
  Reliability and Security Companion (QRS-C)}.\hskip 1em plus 0.5em minus
  0.4em\relax IEEE, 2021, pp. 306--313.

\bibitem{allamanis2019adverse}
M.~Allamanis, ``The adverse effects of code duplication in machine learning
  models of code,'' in \emph{Proceedings of the 2019 ACM SIGPLAN International
  Symposium on New Ideas, New Paradigms, and Reflections on Programming and
  Software}, 2019, pp. 143--153.

\bibitem{boland2012juliet}
T.~Boland and P.~E. Black, ``Juliet 1.1 c/c++ and java test suite,'' \emph{IEEE
  Computer Architecture Letters}, vol.~45, no.~10, pp. 88--90, 2012.

\bibitem{feng2020codebert}
Z.~Feng, D.~Guo, D.~Tang, N.~Duan, X.~Feng, M.~Gong, L.~Shou, B.~Qin, T.~Liu,
  D.~Jiang \emph{et~al.}, ``Codebert: A pre-trained model for programming and
  natural languages,'' \emph{arXiv preprint arXiv:2002.08155}, 2020.

\bibitem{liu2019roberta}
Y.~Liu, M.~Ott, N.~Goyal, J.~Du, M.~Joshi, D.~Chen, O.~Levy, M.~Lewis,
  L.~Zettlemoyer, and V.~Stoyanov, ``Roberta: A robustly optimized bert
  pretraining approach,'' \emph{arXiv preprint arXiv:1907.11692}, 2019.

\bibitem{yao2020assessing}
J.~Yao and M.~Shepperd, ``Assessing software defection prediction performance:
  Why using the matthews correlation coefficient matters,'' \emph{Proceedings
  of the Evaluation and Assessment in Software Engineering}, pp. 120--129,
  2020.

\bibitem{cochran2007sampling}
W.~G. Cochran, \emph{Sampling techniques}.\hskip 1em plus 0.5em minus
  0.4em\relax John Wiley \& Sons, 2007.

\bibitem{cohen1960}
J.~Cohen, ``A coefficient of agreement for nominal scales,'' \emph{Educational
  and psychological measurement}, vol.~20, no.~1, pp. 37--46, 1960.

\bibitem{braun2006using}
V.~Braun and V.~Clarke, ``Using thematic analysis in psychology,''
  \emph{Qualitative research in psychology}, vol.~3, no.~2, pp. 77--101, 2006.

\bibitem{herzig2013impact}
K.~Herzig and A.~Zeller, ``The impact of tangled code changes,'' in \emph{2013
  10th Working Conference on Mining Software Repositories (MSR)}.\hskip 1em
  plus 0.5em minus 0.4em\relax IEEE, 2013, pp. 121--130.

\bibitem{sejfia2021identifying}
A.~Sejfia, Y.~Zhao, and N.~Medvidovi{\'c}, ``Identifying casualty changes in
  software patches,'' in \emph{Proceedings of the 29th ACM Joint Meeting on
  European Software Engineering Conference and Symposium on the Foundations of
  Software Engineering}, 2021, pp. 304--315.

\bibitem{herzig2016impact}
K.~Herzig, S.~Just, and A.~Zeller, ``The impact of tangled code changes on
  defect prediction models,'' \emph{Empirical Software Engineering}, vol.~21,
  no.~2, pp. 303--336, 2016.

\bibitem{herbold2022fine}
S.~Herbold, A.~Trautsch, B.~Ledel, A.~Aghamohammadi, T.~A. Ghaleb, K.~K.
  Chahal, T.~Bossenmaier, B.~Nagaria, P.~Makedonski, M.~N. Ahmadabadi
  \emph{et~al.}, ``A fine-grained data set and analysis of tangling in bug
  fixing commits,'' \emph{Empirical Software Engineering}, vol.~27, no.~6, pp.
  1--49, 2022.

\bibitem{mann1947test}
H.~B. Mann and D.~R. Whitney, ``On a test of whether one of two random
  variables is stochastically larger than the other,'' \emph{The annals of
  mathematical statistics}, pp. 50--60, 1947.

\bibitem{zhao2021impact}
Y.~Zhao, L.~Li, H.~Wang, H.~Cai, T.~F. Bissyand{\'e}, J.~Klein, and J.~Grundy,
  ``On the impact of sample duplication in machine-learning-based android
  malware detection,'' \emph{ACM Transactions on Software Engineering and
  Methodology (TOSEM)}, vol.~30, no.~3, pp. 1--38, 2021.

\bibitem{ain2019systematic}
Q.~U. Ain, W.~H. Butt, M.~W. Anwar, F.~Azam, and B.~Maqbool, ``A systematic
  review on code clone detection,'' \emph{IEEE access}, vol.~7, pp.
  86\,121--86\,144, 2019.

\bibitem{sajnani2016sourcerercc}
H.~Sajnani, V.~Saini, J.~Svajlenko, C.~K. Roy, and C.~V. Lopes, ``Sourcerercc:
  Scaling code clone detection to big-code,'' in \emph{Proceedings of the 38th
  International Conference on Software Engineering}, 2016, pp. 1157--1168.

\bibitem{piantadosi2019fixing}
V.~Piantadosi, S.~Scalabrino, and R.~Oliveto, ``Fixing of security
  vulnerabilities in open source projects: A case study of apache http server
  and apache tomcat,'' in \emph{2019 12th IEEE Conference on software testing,
  validation and verification (ICST)}.\hskip 1em plus 0.5em minus 0.4em\relax
  IEEE, 2019, pp. 68--78.

\bibitem{croft2022investigation}
R.~Croft, M.~A. Babar, and L.~Li, ``An investigation into inconsistency of
  software vulnerability severity across data sources,'' in \emph{2022 IEEE
  International Conference on Software Analysis, Evolution and Reengineering
  (SANER)}.\hskip 1em plus 0.5em minus 0.4em\relax IEEE, 2022, pp. 338--348.

\bibitem{gama2014survey}
J.~Gama, I.~{\v{Z}}liobait{\.e}, A.~Bifet, M.~Pechenizkiy, and A.~Bouchachia,
  ``A survey on concept drift adaptation,'' \emph{ACM computing surveys
  (CSUR)}, vol.~46, no.~4, pp. 1--37, 2014.

\bibitem{le2019automated}
T.~H.~M. Le, B.~Sabir, and M.~A. Babar, ``Automated software vulnerability
  assessment with concept drift,'' in \emph{2019 IEEE/ACM 16th International
  Conference on Mining Software Repositories (MSR)}.\hskip 1em plus 0.5em minus
  0.4em\relax IEEE, 2019, pp. 371--382.

\bibitem{mcintosh2017fix}
S.~McIntosh and Y.~Kamei, ``Are fix-inducing changes a moving target? a
  longitudinal case study of just-in-time defect prediction,'' \emph{IEEE
  Transactions on Software Engineering}, vol.~44, no.~5, pp. 412--428, 2017.

\bibitem{fuglede2004jensen}
B.~Fuglede and F.~Topsoe, ``Jensen-shannon divergence and hilbert space
  embedding,'' in \emph{International Symposium onInformation Theory, 2004.
  ISIT 2004. Proceedings.}\hskip 1em plus 0.5em minus 0.4em\relax IEEE, 2004,
  p.~31.

\bibitem{kendall1938}
M.~G. Kendall, ``A new measure of rank correlation,'' \emph{Biometrika},
  vol.~30, no. 1/2, pp. 81--93, 1938.

\bibitem{sawadogo2022sspcatcher}
A.~D. Sawadogo, T.~F. Bissyand{\'e}, N.~Moha, K.~Allix, J.~Klein, L.~Li, and
  Y.~Le~Traon, ``Sspcatcher: Learning to catch security patches,''
  \emph{Empirical Software Engineering}, vol.~27, no.~6, pp. 1--32, 2022.

\bibitem{garg2022learning}
A.~Garg, R.~Degiovanni, M.~Jimenez, M.~Cordy, M.~Papadakis, and Y.~LeTraon,
  ``Learning from what we know: How to perform vulnerability prediction using
  noisy historical data,'' \emph{arXiv preprint arXiv:2207.11018}, 2022.

\end{thebibliography}

\end{document}